\documentclass{mn2e}
\usepackage{psfig}

\def \kmsMpc{\hbox{kms}^{-1}\,\hbox{Mpc}^{-1}}
\def \hkpc{\,h^{-1} \hbox{kpc}}
\def \hMpc{\,h^{-1} \hbox{Mpc}}
\def \hMsol{\,h^{-1} \hbox{M}_{\odot}}

\def \h2gcm3{h^{2} \, \hbox{g}\,\hbox{cm}^{-3}}

\title{The impact of galaxy formation on X-ray groups}
\author[S.T. Kay et al.]
       {Scott T. Kay,$^{1}$\thanks{E-mail: s.t.kay@sussex.ac.uk}
	Peter A. Thomas$^{1}$ and Tom Theuns$^{2,3,4}$ \\
        $^{1}$ Astronomy Centre, University of Sussex, Falmer, 
	Brighton BN1 9QJ\\
	$^{2}$ Institute for Computational Cosmology, Physics Department, 
        University of Durham, South Road, Durham DH1 3LE\\
	$^{3}$ Institute of Astronomy,  Madingley Road, Cambridge CB3 0HA\\
	$^{4}$ Universitaire Instelling Antwerpen, Universiteitsplein 1, B-2610 
	Antwerpen, Belgium.
	}

\begin{document}

\maketitle

\label{firstpage}

\begin{abstract}
Using hydrodynamical simulations of the $\Lambda$CDM cosmology,
that include both radiative cooling and a phenomenological model
for star formation and supernovae feedback, we investigate the impact 
of galaxy formation on the X-ray properties of groups at redshift zero. 
Motivated by the observed ``break'' in the $L_{\rm x}-T_{\rm x}$ relation at
$kT_{\rm x} \sim 1-2$keV, our feedback model is based on the assumption that 
supernovae imprint a temperature scale on the hot gas, with the star formation
rate and corresponding reheated gas mass then depending only on the available 
energy budget. We demonstrate that a strong feedback model with a heating 
temperature comparable to this break ($kT_{\rm SN}=2$keV), and an energy budget 
twice that available from supernovae ($\epsilon=2$), raises the core entropy 
of groups sufficiently to produce an adequate match to their observed X-ray 
properties. A lower value of $\epsilon$ increases the star formation
rate without significantly affecting the X-ray properties of groups, and a model 
with $\epsilon \sim 0.1$ reproduces the observed fraction of baryons in stars.
However, a heating temperature which is lower than the virial temperatures of 
the groups leads to an excess of cooling gas that boosts their X-ray 
luminosities, due to the failure of the reheated material to escape from the 
gravitational potential. A limited study of numerical resolution effects
reveals that the temperature of poorly-resolved objects is underestimated,
therefore (in our case) a fully-resolved group population would lead to
a steeper $L_{\rm x}-T_{\rm x}$ relation, bringing our results into even
better agreement with the observations.
\end{abstract}


\section{Introduction}
\label{sec:intro}

Groups 
\footnote{We use the term ``groups'' to mean systems 
with X-ray temperatures of order 1 keV or lower}
are the most common agglomerates of galaxies in the
low-redshift universe and also harbour the majority of currently detected
baryons (Fukugita, Hogan \& Peebles 1998), observed because of
their emission of soft X-rays ($kT \sim$ 0.1-1 keV). The simplest
explanation for the properties of this gas is that it was
shock-heated during the gravitational collapse and virialization
of the system. For self-similar ($\Omega=1$) systems, this leads to a 
simple scaling between X-ray luminosity and temperature, 
$L_{\rm x} \propto T^{3/2} (1+z)^{3/2} \Lambda(T)$, 
where $\Lambda(T)$ denotes the cooling rate of the gas.

In the largest clusters, the virial temperature is sufficiently 
high that the gas predominantly emits bremsstrahlung radiation, leading to
the scaling $L_{\rm x} \propto T^2$ (Kaiser 1986). This result was 
recently vindicated by Allen, Schmidt \& Fabian (2002) using high-quality 
{\it Chandra} data. In smaller systems, the increasing importance of line 
emission boosts the luminosity and so should flatten the $L_{\rm x}-T_{\rm x}$ 
relation. Observationally the converse is true, with the slope of the relation 
for low-mass groups being as high as $\sim 6$ (e.g. see Xue \& Wu 2000 for a 
recent compilation). This result is also reflected in the low amplitude of the 
soft X-ray background, around an order of magnitude lower than predicted from 
gravitational heating models (e.g. Pen 1999; Bryan \& Voit 2001; 
Wu, Fabian \& Nulsen 2001).

Evrard \& Henry (1991; see also Kaiser 1991; Bower 1997; Voit et al. 2002) 
argued that the deficit in luminosity is due to additional physical processes 
raising the {\it entropy} of the gas. (Entropy is invariant under adiabatic 
processes, making it a useful variable for characterizing the effects 
of cooling and heating on the gas distribution.) Gravitational heating 
leads to the mean entropy of a self-similar system scaling as 
$s \propto T/(1+z)^2$
\footnote{We use the term {\it entropy} to mean $s=T/\rho^{\gamma-1}$,
and assume $\gamma=5/3$, for a monatomic ideal gas}. Increasing the entropy 
of the gas corresponds to a hotter but less-concentrated distribution. 
Consequently, for a given increase in entropy, the X-ray emissivity is 
lowered preferentially in low-mass systems, 
steepening the $L_{\rm x}-T_{\rm x}$ relation.
Ponman, Cannon \& Navarro (1999) measured the entropy of groups and clusters at 
10 per cent of their virial radii, and showed that low-mass clusters and 
groups deviated from the self-similar prediction, tending to a constant value of 
$\sim 100 h^{-1/3}$ keV cm$^2$ (albeit with very large scatter).

The precise origin of the excess entropy is still unclear, but is
most probably a direct consequence of the galaxy formation process
(Evrard \& Henry 1991). For example, galactic winds from 
supernovae explosions and/or active galactic nuclei (AGN) could heat the gas, 
raise its entropy, and allow it to convect out to lower density environments. 
Winds have already been observed in low-redshift starburst (e.g. Heckman 2000) and 
high-redshift Lyman Break (e.g. Pettini et al. 2001) galaxies, and the 
presence of metal-enriched gas in low-density regions suggests
galactic material has been expelled from haloes (e.g. Schaye et al. 2000).
Note that in hierarchical models, at least some heating (feedback) is required
in order to regulate the star formation rate (e.g. White \& Rees 1978; 
Balogh et al. 2001). 

The amount of {\it energy} required to generate excess entropy depends on 
environment. If the gas was close to mean density when heated (the 
{\it pre-heating} scenario) only $\sim 0.3$keV of energy per particle is 
required (Lloyd-Davies, Ponman \& Cannon 2000). More realistically, if the 
heating occurred in high-density regions,
as much as 1-3 keV per particle is required (e.g. Wu, Fabian \& Nulsen 2000; 
Kravtsov \& Yepes 2000; Bower et al. 2001; Borgani et al. 2002). Note, however 
that the total energy production is smaller in the latter case as only gas in the 
cores of haloes need be heated.

Excess entropy is also generated by the process of star/black-hole formation itself, 
removing cold, low-entropy material and causing higher-entropy material to 
flow in to replace it (Knight \& Ponman 1997). Models using only radiative cooling
to generate the excess entropy have been implemented by a number of 
authors, and have proved successful in reproducing a number of X-ray scaling relations 
(e.g. Pearce et al. 2000; Bryan 2000; Voit \& Bryan 2001; Muanwong et al. 2001,2002; 
Wu and Xue 2002a,2002b; Dav\'{e} et al. 2002). Voit \& Bryan (2001) 
demonstrated that the entropy--temperature locus of gas with a range
of temperatures and densities set by equating their cooling times 
to the age of the universe, closely matches the relation observed by 
Ponman et al. (1999). Voit \& Bryan further argued for the presence of an 
entropy {\it floor} by asserting that cooled gas will either form stars or be 
heated (by stars or AGN) and move to a higher adiabat.

The aim of this paper is to investigate how galaxies which form in a 
cosmological simulation impact upon the X-ray properties of groups. This
treatment allows us to incorporate cooling and heating processes in
a self-consistent manner. Specifically, we include
the effects of both radiative cooling/star formation and feedback of energy 
from supernovae in our simulations, but neglect an AGN component (although
in broad terms, the effects of including AGN may largely mimic the
desired effects of supernovae, since both are ultimately powered by cooling
gas). Motivated by the observed ``break'' in the $L_{\rm x}-T_{\rm x}$ relation 
at the interface between groups and clusters ($kT \sim 1-2$keV), our feedback
model is fundamentally based on the assumption that cold gas is reheated to a 
fixed temperature and the mass ratio of reheated gas to stars is determined
by the overall energy budget. As we shall see, our model is able to
reproduce the required entropy excess, provided that the heating temperature
is sufficiently high to allow the majority of reheated gas to escape from
haloes, while varying the energy budget primarily controls the fraction
of baryons in stars.

We organize the remainder of this paper as follows. In 
Section~\ref{sec:method} we outline our method, giving details of the 
simulations as well as our approach to cooling, star formation and feedback.
Section~\ref{sec:results} presents the predicted X-ray group scaling 
relations for our fiducial simulation at $z=0$. In Section~\ref{sec:model} 
we investigate our feedback model in more detail, concentrating on the effects of 
varying parameters. In Section~\ref{sec:resolution}, we perform a limited 
study of numerical resolution effects and summarize our conclusions in 
Section~\ref{sec:conclusions}.

\section{Simulation details}
\label{sec:method}

\subsection{Overview}

Results are presented for the currently-favoured $\Lambda$CDM 
cosmology, setting the density parameter, $\Omega_m=0.35$, 
Hubble constant, $h=H_0/100 \kmsMpc =0.71$, cosmological constant, 
$\Omega_{\Lambda}=\Lambda/3H_0^2=0.65$, baryon density, 
$\Omega_b = 0.038$, and power spectrum normalisation, 
$\sigma_8=0.7$. The latter value, recently measured from both
cluster abundances (e.g. Seljak 2002) and galaxy clustering 
(Lahav et al. 2002), is around 20 per cent lower than previous determinations 
which deduced $\sigma_8 \sim 0.9$ (although, weak lensing statistics 
still prefer the higher value; see e.g. Refregier et al. 2002). We find
that increasing $\sigma_8$ from 0.7 to 0.9 has only a small effect
on the results presented in this paper, but the lower value gives us a 
numerical advantage, requiring fewer timesteps to evolve to $z=0$.

Simulation data were generated using a parallel (OpenMP) version
of the {\sc hydra} code (Couchman, Thomas \& Pearce 1995;
Theuns et al. 1998), a combination of the AP$^3$M algorithm 
(Couchman 1991) to compute gravitational forces and Smoothed 
Particle Hydrodynamics (SPH; see Monaghan 1992) to calculate 
gas forces. For the gas, neighbour lists were computed using the
recursive binning technique described in Theuns et al. (1998), 
allowing the density to be estimated for gas particles in 
low-density regions. SPH forces were calculated using an artificial
viscosity based on pairwise particle separations (Monaghan 1992) 
rather than the local velocity divergence. Our SPH implementation
now also includes the force-correction terms suggested by Springel \&
Hernquist (2002), resulting in improved entropy conservation.

Initial conditions were generated using the {\sc cosmic} package 
supplied with the publicly available version of {\sc hydra}. We
made a small modification to {\sc cosmic} so that the large-scale 
structure is preserved when increasing the numerical resolution. 
For our main results, we adopted a comoving box-size 
of $50 \hMpc$, populated with a perturbed grid of 
2,097,152 ($128^3$) dark matter and gas particles each. 
(Note that the box-size is big enough to give statistically
meaningful results; Bryan \& Voit 2001.)
These choices set our gas and dark matter particle masses to  
$m_{\rm gas}=6.3\times 10^{8}\hMsol$ 
and $m_{\rm dark}=5.2\times 10^{9}\hMsol$ respectively.
We have chosen this resolution so not to expect significant
heating of the gas due to two-body encounters with the dark matter
particles (Steinmetz \& White 1997). 
Gravitational forces were softened on short scales using a 
spline kernel with an equivalent Plummer softening length
of $\epsilon_{\rm p} \equiv \epsilon_{\rm s}/2.34 = 10 \hkpc$. 
Gas forces were also limited by fixing the minimum 
SPH smoothing radius to the gravitational softening 
$2h_{\rm min}=\epsilon_{\rm s}$.

All but the smallest runs were performed on the UK 
Cosmology Consortium COSMOS Origin 3800 supercomputer in Cambridge,
using between 10 and 20 processors. Simulations were started at $z=49$ 
and evolved to $z=0$, typically taking between 2000 and 4500 steps, 
depending on the choice of resolution and feedback model.

\subsection{Radiative cooling and star formation}

Radiative cooling was implemented following the method
of Thomas \& Couchman (1992), using tabulated cooling
rates given by Sutherland \& Dopita (1993). For the present paper, 
we adopt a global metallicity scaling law, $Z=0.3(t/t_0)Z_{\odot}$, to 
approximate the gradual enrichment of the intergalactic medium. In 
principle, we could model metal enrichment self-consistently, since 
it is a by-product of the supernovae explosions. (We have already 
done this to study the impact of dwarf galaxies on the high-redshift 
intergalactic medium; Theuns et al. 2002.) In order to simplify matters 
however, we have decided to leave such a study to a subsequent paper.

We follow the star formation prescription used by Kay et al.(2002, 
hereafter K2002), converting gas with $\delta>5000$ and $T<50,000$K 
into stars at the end of each timestep. This density threshold is low 
compared to measured empirical density thresholds in galaxies 
(e.g. Kennicutt 1998) but is close to our resolved density maximum. 
Increasing the density threshold above the resolution threshold induces
a delay to the star formation, increasing the gas to stars ratio within 
each galaxy (K2002). 

Subsequent to the star formation routine, we identify groups of 10 stars 
within the current gravitational softening length and merge them into 
a single {\it galaxy fragment}. These fragments are comparable in
mass to the dark matter particles and the reduction in the number
of particles provides a modest decrease in the CPU time per step.

\subsection{Feedback}

The essence of any feedback model is to create a diffuse hot phase 
which powers a wind, providing a mechanism 
to transport gas away from star forming regions. However, explicitly 
modelling a multi-phase medium in a cosmological simulation is impossible 
because of its limited resolution. This problem has prompted two
approaches to implementing feedback within such simulations.

One method is to assume that each resolution element (in the case of SPH, a 
single gas particle) is representative of a multi-phase medium (cold and hot gas
and stars), and to solve explicitly for the flux of material between each 
phase, using a  set of physical arguments as laid out by McKee \& Ostriker (1977). 
This approach has been implemented and proved effective by various authors, 
using both grid-based codes (e.g. Yepes et al. 1997) and SPH 
(e.g. Hultman \& Pharasyn 1999; Springel \& Hernquist 2003). The main 
advantage of this method is that it does not depend explicitly on 
numerical resolution, and so should be less sensitive to discreteness 
effects. On the other hand, the phases cannot move independently (although
see Springel \& Hernquist 2003).

An alternative approach, which we take here, is to allow each
particle to represent only one phase and use a set of simple rules
to overcome any associated discreteness problems.
This has the advantage of being much simpler to implement than the 
previous method and allows each phase to move freely. 
K2002 reviewed a number of feedback models within this context; in this 
paper, we consider an improved version of the thermal feedback approach 
studied by K2002 and outline details of this method below.

\subsubsection{Feedback energetics}

\begin{figure}
\centering
\centerline{\psfig{file=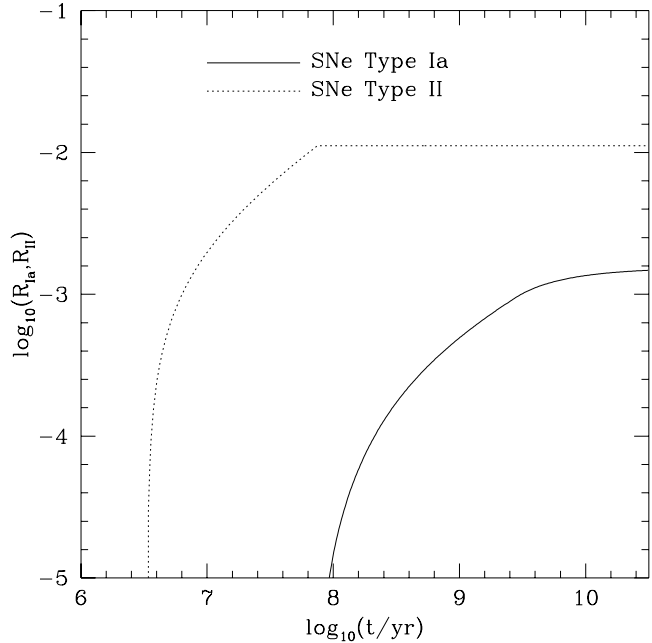,height=9cm}}
\caption{Cumulative number of supernovae per solar mass of
stars formed versus time since the star formation episode,
using power-law fits of Lia et al. 2002 and
assuming a Salpeter IMF.}
\label{fig:sne}
\end{figure}

The amount of energy available from each star particle at time,
$t$, to heat the gas over a timestep, $\Delta t$, is 
\begin{equation}
E_{\rm SN} = 10^{51} \, \epsilon \, 
\left( \frac{m_*}{M_{\odot}} \right) 
\int_{t-t_0}^{t+\Delta t-t_0} \, 
\left[ r_{\rm Ia}(\tau)/2 + r_{\rm II}(\tau) \right]
\, \, d\tau \, \, {\rm erg},
\label{eqn:esn}
\end{equation}
where $m_*$ is the mass of a star particle, 
$t_0$ is the star formation time and $r_{\rm Ia}$ \& $r_{\rm II}$ are the
rates of Type~Ia and Type~II supernovae (SNe) per solar mass of stars formed.
(We assume each Type~Ia supernova contributes half the energy of a Type~II 
supernova.)

Lia et al. (2002) tabulated $r(\tau)$ for various initial mass functions (IMF), 
using the Padova stellar evolutionary tracks (Portinari et al. 1998). We
adopt the Lia et al. rates for a Salpeter IMF (given in their Table~2)
and illustrate integrated versions in Fig.~\ref{fig:sne}. 
The contribution from Type~II SNe rises dramatically after 
approximately 3 million years, lasting for around 70 million years. 
Type~Ia SNe start contributing later, at around 100 million years, but are 
still going off after 10 billion years. The total number of Type~II SNe is 
around an order of magnitude higher than the number of Type~Ia SNe, and so 
provides the dominant source of feedback of energy to the interstellar 
medium. 

The parameter $\epsilon$ modulates the overall energy available from 
the SNe, which we term the feedback efficiency. Thornton et al. (1998) 
suggested that the maximal value ($\epsilon=1$) may be at least a factor 
of 5 too high once radiative losses are accounted for. 
As we shall see in Section~\ref{sec:model}, our model does not
necessarily require a high value of $\epsilon$ to generate
the required excess entropy; varying this parameter primarily 
alters the fraction of baryons in galaxies. 

\subsubsection{Feedback implementation}

The thermal feedback model studied by K2002 is effective in
reducing the masses/luminosities of galaxies but fails to
reproduce the X-ray scaling relations of groups. In their model,
each star particle distributes the energy available at each
timestep (equation~\ref{eqn:esn}) over the surrounding gas.
For equal mass particles, the maximum amount of energy a gas particle 
can receive from a single star particle is enough to heat it by 
$\sim 2$keV (for $\epsilon=1$). In practice, gas particles escape 
before receiving this amount
of energy and the result is an average heating temperature around
an order of magnitude lower. Much of this gas then fails to escape 
from group and cluster-sized haloes with comparable or higher
virial temperatures, and instead re-cools, boosting their X-ray luminosities
(we will demonstrate this effect in Section~\ref{sec:model}).

We have devised an alternative method with the clear intention of
heating the gas to a higher temperature without necessarily increasing
the available energy. In the context of X-ray scaling relations, our 
new model is motivated by the presence of a ``break'' in the 
$L_{\rm x}-T_{\rm x}$ relation at $kT_{\rm x} \sim 1-2 $keV; such a feature may 
be an imprint of a temperature scale introduced by the heating mechanism.
We base the model on the physical assumption that as more and more 
supernovae go off, it is not the temperature of the gas that is 
progressively increased (as in the previous method), but the mass fraction 
of material converted from the cold phase to the hot phase.  
To implement such a model, we equate the available energy from a mass of 
newly formed stars, $m_*$ to the energy required to raise the temperature of 
a mass of cold gas, $\delta m_{\rm gas}$ by a fixed amount, $T_{\rm SN}$
\begin{equation}
\delta m_{\rm gas} \, (3 kT_{\rm SN}/2 \mu m_{\rm H}) = E_{\rm SN},
\label{eqn:energy}
\end{equation}
where $E_{\rm SN}$ is given in equation~\ref{eqn:esn} and we set 
$\mu m_{\rm H} = 10^{-24}$g, 
the mean molecular weight of a gas with $Y=0.24,Z=0$. In practice, 
$T_{\rm SN}$ is much greater than the temperature of the cold gas 
($\sim 10^{4}$K) so the heating introduces a temperature scale to 
the hot gas.

Assuming now that $m_*$ is the mass of a 
star particle, the fractional mass of a gas particle
capable of being heated to $T_{\rm SN}$ is
\begin{equation}
f_{\rm heat} = {\delta m_{\rm gas} \over m_{\rm gas}}
= 0.209 \, \epsilon 
\, \left( \frac{R(t-t_0,\Delta t)}{0.001} \right)
\, \left( \frac{kT_{\rm SN}}{{\rm keV}} \right)^{-1}
,
\label{eqn:fheat}
\end{equation}
using the fact that $m_{*}=m_{\rm gas}$ in our method
and $R(t-t_0,\Delta t)$ is the integral on the right hand side of
equation~\ref{eqn:esn}. (For a typical timestep, 
$\Delta t \sim 5 \times 10^{6}$yr, $\left< R \right> \sim 0.001$.)
Note that in K2002's model, $f_{\rm heat}=1$ and 
$kT_{\rm SN}$ is calculated, the opposite of our model.
For every star particle, we calculate the radius which encloses
32 gas particles, and (since generally $f_{\rm heat}<1$
\footnote{Exceptions to this rule can occur if 
$\epsilon/kT_{\rm SN}>5$; this is true for only one model considered
in this paper, where $\epsilon=1$ \& $kT_{\rm SN}=0.1$keV. As we shall 
see however, this particular model fails because $kT_{\rm SN}$
is too low, not because less gas is reheated.}
) apply 
equation~\ref{eqn:fheat} to the first gas particle within
this radius with $T<50,000$K. 
Since each gas particle represents only one phase in our 
simulations, we can only heat the whole particle to a temperature
$T_{\rm SN}$ or not heat it at all. To mimic the effect of
heating a fraction of the gas, we draw a random number for each particle, 
$0\le r \le 1$, and heat it to $T_{\rm SN}$ if $f_{\rm heat}<r$.

For our fiducial model, for which we present results in the
next section, we set $\epsilon=2$ \& $kT_{\rm SN}=2$keV,
twice the maximum available energy from standard supernovae calculations,
but in reasonable agreement with that suggested in previous studies
(e.g. Kravtsov \& Yepes 2000; Wu, Fabian \& Nulsen 2001;
Bower et al. 2001; Borgani et al. 2002). As we will show, this model
predicts X-ray properties of groups in good agreement with the data.
We will demonstrate in Section~\ref{sec:model} however, that 
a good match to the scaling relations is reasonably insensitive to this choice of 
$\epsilon$, but does require the heating temperature to be sufficiently
high in order for gas to escape from the gravitational potential of the 
halo. The fiducial temperature, which is significantly larger 
than typically found when using K2002's model, also alleviates the need for 
preventing gas from cooling radiatively over a fixed period of time (as 
implemented by K2002), because cooling gas only starts to 
become thermally unstable at lower temperatures ($kT \sim 0.2-1$keV for 
$Z=0-0.3Z_{\odot}$). 


\section{X-ray properties of groups in the fiducial model}
\label{sec:results}

We begin by demonstrating that our fiducial feedback model
provides an adequate match to current observational data at $z=0$, 
given the level of uncertainty in these measurements. 

Our group catalogues are constructed using the procedure outlined by
Muanwong et al. (2002, hereafter M2002), where full details may be found.
To summarize, spheres of baryon and dark matter particles are located
with constant mean internal overdensity, $\Delta$, relative to the comoving 
critical density. For our chosen cosmology, $\Delta=111$ is approximately the 
predicted value by the spherical top-hat collapse model for a virialized sphere 
(Eke, Navarro \& Frenk 1998). Unless specified, we set $\Delta=200$ for our 
main results but note that the X-ray properties are insensitive to using 
$\Delta=200$ rather than $\Delta=111$. Only objects containing at least 500 
baryon plus 500 dark matter particles are retained, setting a minimum halo mass of 
$2.9 \times 10^{12} h^{-1}$M$_{\odot}$. At $z=0$, our fiducial catalogue contains
172 objects.

Once the catalogues were generated, it was found that a small fraction
($\sim 5$ per cent) of groups have X-ray temperatures very much in excess 
of the average 
for their mass (i.e. there are hot outliers on the temperature-mass relation). 
On inspection, these objects are infalling sub-clumps on the outskirts of 
larger groups, but have hot gas which has been shocked to the temperature of 
the more massive object. We choose to eliminate these objects by setting
the requirement that each group in the catalogue has an X-ray temperature 
less than twice the mean value for its mass.

\subsection{Entropy}

\begin{figure}
\centering
\centerline{\psfig{file=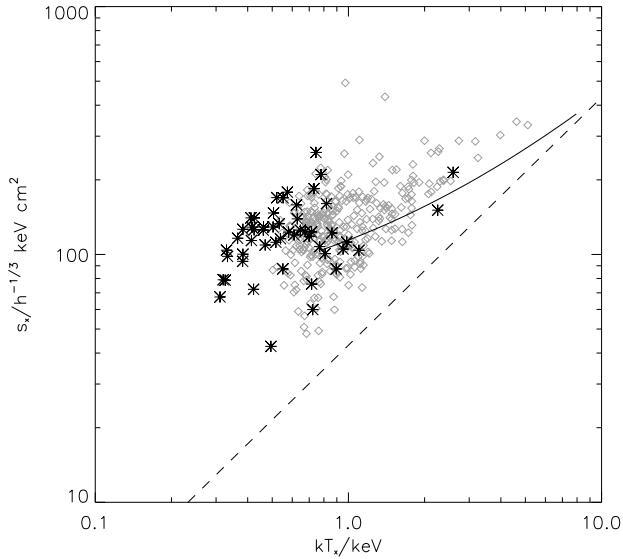,height=8cm}}
\caption{Core entropy (measured at $0.1 r_{\rm vir}$) against X-ray 
temperature. Stars represent groups in the fiducial model and diamonds
from the {\it Radiative} simulation in M2002. The solid line represents the trend
observed by Ponman et al. (1999) and the dashed line is a fit
to the self-similar relation obtained from {\it Non-radiative} simulations.}
\label{fig:entemrel}
\end{figure}

The effect of the removal of hot gas from a halo on its X-ray properties
can best be characterized by the change in entropy of the gas. 
As the hot gas is depleted, higher entropy material flows inwards
and is adiabatically heated
(i.e. it develops a hotter, less concentrated distribution)
in order to maintain gravitational support. This leads to a rise in
the X-ray temperature of the gas but a {\it decrease} in its X-ray luminosity
(Knight \& Ponman 1997; Pearce et al. 2000).
Ponman et al. (1999) demonstrated that lower temperature objects have shallower 
surface brightness profiles and consequently larger excess entropies, relative
to the expected self-similar relation for a model neglecting galaxy formation
(we refer to this as a {\it Non-radiative} model). 

Following Ponman et al., we measure the entropy of each object at
0.1$r_{\rm vir}$ as a function of X-ray temperature. The latter is 
evaluated within a soft--band of energy range 0.3-1.5 keV
\begin{equation}
T_{\rm x} = {\sum_i \, m_i \rho_i \Lambda_{\rm soft}(T_i,Z) T_i
             \over
             \sum_i \, m_i \rho_i \Lambda_{\rm soft}(T_i,Z)},
\label{eqn:txsoft}
\end{equation}
where the sum is performed over all hot gas particles ($T>10^{5}$K)
within the virial radius of the group, each with mass, $m_i$, density, 
$\rho_i$ and temperature, $T_i$. The soft-band cooling function, 
$\Lambda_{\rm soft}$ is calculated using the Raymond \& Smith (1977) models. 
We then estimate the entropy of each group
\begin{equation}
s_{\rm x}=kT_{\rm x}/\left<n_{\rm e}\right>^{2/3},
\label{eqn:entdef}
\end{equation}
where 
\begin{equation}
\left<n_{\rm e}\right>={\eta \over m_{\rm H}} \, 
                       {3M \over 4\pi r^2 \delta r}
\label{eqn:n_e}
\end{equation}
is the electron density, averaged over all hot gas particles (of total mass,
$M$) within a shell of width, $\delta r = 0.1r_{\rm vir}$, centered on 
$r=0.1r_{\rm vir}$. 
Only groups containing 10 or more particles within this region are plotted.
We set the number of electrons per baryon, $\eta=0.88$, appropriate
for a primordial gas ($Y=0.24,Z=0$) although insensitive to the assumed range of
$Z$.

Fig.~\ref{fig:entemrel} shows the simulated entropy--temperature
relation (stars), as well as groups/clusters from the {\it Radiative} 
simulation studied by M2002 (diamonds). Our fiducial model agrees well
with M2002's result where the two samples overlap, as well as with the general 
trend observed by Ponman et al. (solid curve), clearly demonstrating an entropy in 
excess of the self-similar prediction ($s_{\rm x} \propto T_{\rm x}$, dashed line). 
The increasing impact of the galaxies with decreasing halo mass leads to an increase
in the excess entropy, consistent with the expectation of an entropy floor at
$s_{\rm x} \sim 100 \, h^{-1/3}$ keV cm$^2$.

\subsection{X-ray temperature}

\begin{figure}
\centering
\centerline{\psfig{file=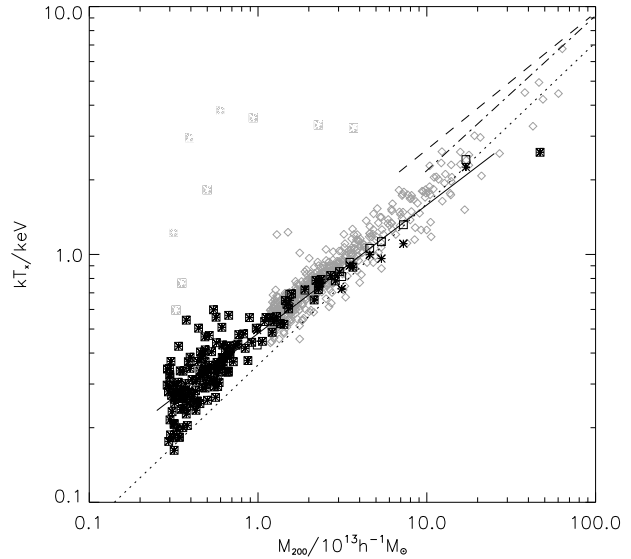,height=8cm}}
\caption{X-ray temperature--mass relation for the fiducial model
(stars). Temperatures calculated when excluding particles 
from within the cooling radius are shown as squares.
Also plotted is the M2002 {\it Radiative} sample (diamonds). 
The dotted line is the relation
between virial temperature and mass. Dashed and dot--dashed lines represent
best-fit relations to the observational data presented in 
Xu, Jin \& Wu (2001) for NFW \& $\beta$ model surface brightness 
profiles respectively.}
\label{fig:tm}
\end{figure}

Increasing the entropy of the gas leads to a higher X-ray temperature
at fixed mass, resulting in an increase in the normalisation of the 
temperature--mass relation (e.g. Muanwong et al. 2001; Thomas et al. 2002).
We plot the fiducial temperature--mass relation in Fig.~\ref{fig:tm} (stars),
with the solid line representing a least--squares fit to the data.
The square points show the result of excluding emission from within the 
cooling radius\footnote{The cooling radius is defined as that which
encloses hot gas with an average cooling time equal to 6 billion years}
of each object. This has the effect of slightly increasing the temperatures 
of some of the larger objects, without significantly affecting the 
slope or normalisation of the relation.

The result is in reasonable agreement with the {\it Radiative} model 
studied by M2002 (diamonds), but predicts a flatter relation
($T_{\rm x} \propto M_{200}^{0.5}$ as opposed to 
 $T_{\rm x} \propto M_{200}^{0.6}$). Lower-mass groups have X-ray 
temperatures in excess of their virial temperatures 
($T_{\rm vir} \propto M_{200}^{2/3}$, dotted line), as expected from
the increase in entropy of these systems. In the highest mass group, 
the presence of cooler reheated gas within the virial radius biases
its X-ray temperature lower than the virial temperature. It is possible 
that a good fit to higher-mass systems will require an even higher
heating temperature than employed here.

The dashed and dot--dashed lines in Fig.~\ref{fig:tm} represent power-law fits
to the observed temperature--mass relation for clusters, compiled by 
Xu, Jin \& Wu (2001), using NFW and $\beta$ model surface brightness profiles 
respectively. Both fits
are offset from the simulated relation by up to 40 per cent. A likely reason
for this offset is that the masses are underestimated when extrapolating 
X-ray surface brightness profiles to the virial radius. M2002 demonstrated 
that by estimating cluster masses this way (assuming a relation between 
the observable extent of an object and its true size) the resulting 
temperature--mass relation was in much closer agreement with the observations.
This conclusion was strengthened when the same dataset was shown to be in
good agreement with the temperature--mass relation for a sample of high-resolution 
clusters observed with the {\it Chandra} satellite (Allen et al. 2001), for
masses measured within a density contrast of 2500 (Thomas et al. 2002).

\begin{figure}
\centering
\centerline{\psfig{file=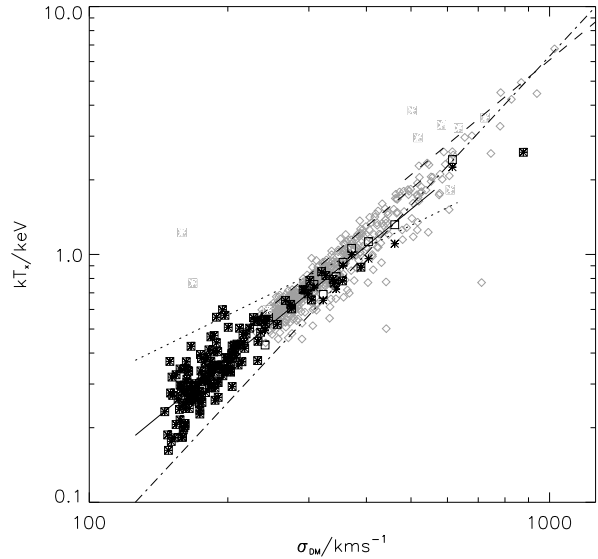,height=8cm}}
\caption{X-ray temperature against dark matter velocity dispersion
for groups in the fiducial model (stars). Square points are for
when emission is excluded from within the cooling radius
(with the solid line representing a least-squares fit to these
objects). Diamonds are from the {\it Radiative} simulation studied by M2002. 
The dot-dashed line is the expected relation when the 
gas and dark matter have equal specific energies ($T \propto \sigma^2$). Dotted 
and dashed lines represent best-fits to the group data in Helsdon \& Ponman (2000) 
and the compiled group and cluster data in Xue \& Wu (2000). 
}
\label{fig:tsigma}
\end{figure}

For groups, the problem of estimating masses from the X-ray data is exacerbated by 
their lower surface brightness, and so galaxy velocity dispersion 
is commonly used as an independent measure of the dynamical size of the halo from 
its temperature. We plot the relation between X-ray temperature and dark matter 
velocity dispersion, $\sigma_{\rm DM}$, for our fiducial model in 
Fig.~\ref{fig:tsigma}. 
(Numerical resolution limits the number of galaxies per halo to a few in most cases, 
preventing us from reliably estimating the galaxy velocity dispersion. Previous
simulations have shown that the galaxies trace the potential reasonably well,
see e.g. Springel et al. 2001; Dav\'{e} et al. 2002.)

\begin{figure}
\centering
\centerline{\psfig{file=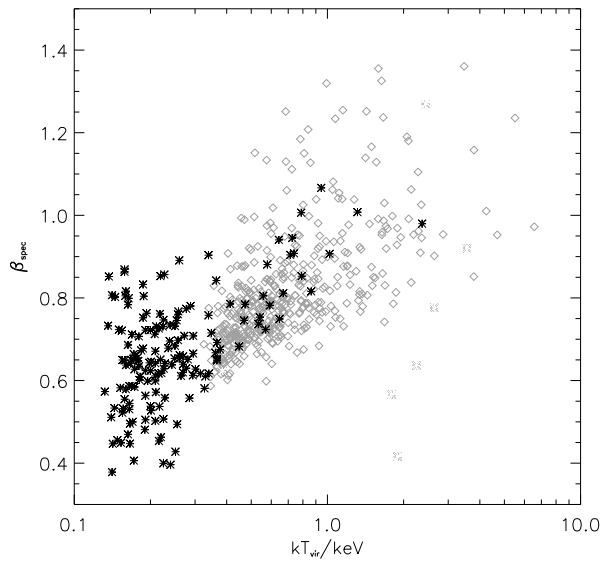,height=8cm}}
\caption{The ratio between X-ray temperature and dark matter
velocity dispersion, $\beta_{\rm spec}$, versus virial temperature
for groups in the fiducial model (stars). Diamonds are from the
{\it Radiative} simulation studied by M2002.}
\label{fig:betaspec}
\end{figure}
The dot-dashed line shows the expected relation if the gas contained the same 
specific energy as the dark matter, i.e. $T_{\rm x} \propto \sigma_{\rm DM}^2$. 
Again, the larger X-ray temperatures of the smaller groups flattens this relation, 
resulting in $T_{\rm x} \propto \sigma_{\rm DM}^{1.5}$ (solid line). To quantify 
this effect, 
we plot the ratio of specific energies between the hot gas and the dark matter
\begin{equation}
\beta_{\rm spec} = {\mu m_{\rm H} \sigma_{\rm DM}^2 \over kT_{\rm x}},
\label{eqn:betaspec}
\end{equation}
in Fig.~\ref{fig:betaspec}.  The smaller objects contain, on average,
X-ray luminous gas that is nearly twice as hot as the dark matter. For the 
higher-mass objects ($kT_{\rm vir} > 1$keV) however, the two components are 
approximately at the same temperature.

Also plotted in Fig~\ref{fig:tsigma} are the best-fits to the group data
presented in Helsdon \& Ponman (2000; dotted line), and the group
and cluster data compiled by Xue \& Wu (2000; dashed lines). The Helsdon \&
Ponman relation is much flatter than the Xue \& Wu line, due to
their sample containing several low--temperature groups with low
velocity dispersions. Removing these outliers from the sample would result in
closer agreement with Xue \& Wu's relation. Our simulated relation
reproduces Xue \& Wu's observational determination,
$T_{\rm x} \propto \sigma_{\rm gal}^{1.56}$, reasonably well.

\subsection{X-ray luminosity}

The increase in entropy decreases the X-ray luminosity of groups.
Since this has a bigger effect on less massive objects, it leads
to a steepening of the $L_{\rm x}-T_{\rm x}$ and $L_{\rm x}-\sigma$ relations,
which we will quantify here. 

As in M2002 we estimate bolometric X-ray luminosity by applying a correction 
to the soft--band emissivity
\begin{equation}
L_{\rm x} = {\Lambda_{\rm bol}(T_{\rm x}) \over \Lambda_{\rm soft}(T_{\rm x})}
        \,  \sum_i {m_i \rho_i \Lambda_{\rm soft}(T_i,Z) \over
            (\mu m_{\rm H})^2},
\label{eqn:lx}
\end{equation}
where $T_{\rm x}$ is calculated using equation~\ref{eqn:txsoft},
$\Lambda_{\rm bol}$ is the bolometric cooling function used in the
simulations (see Section~2).  

\begin{figure}
\centering
\centerline{\psfig{file=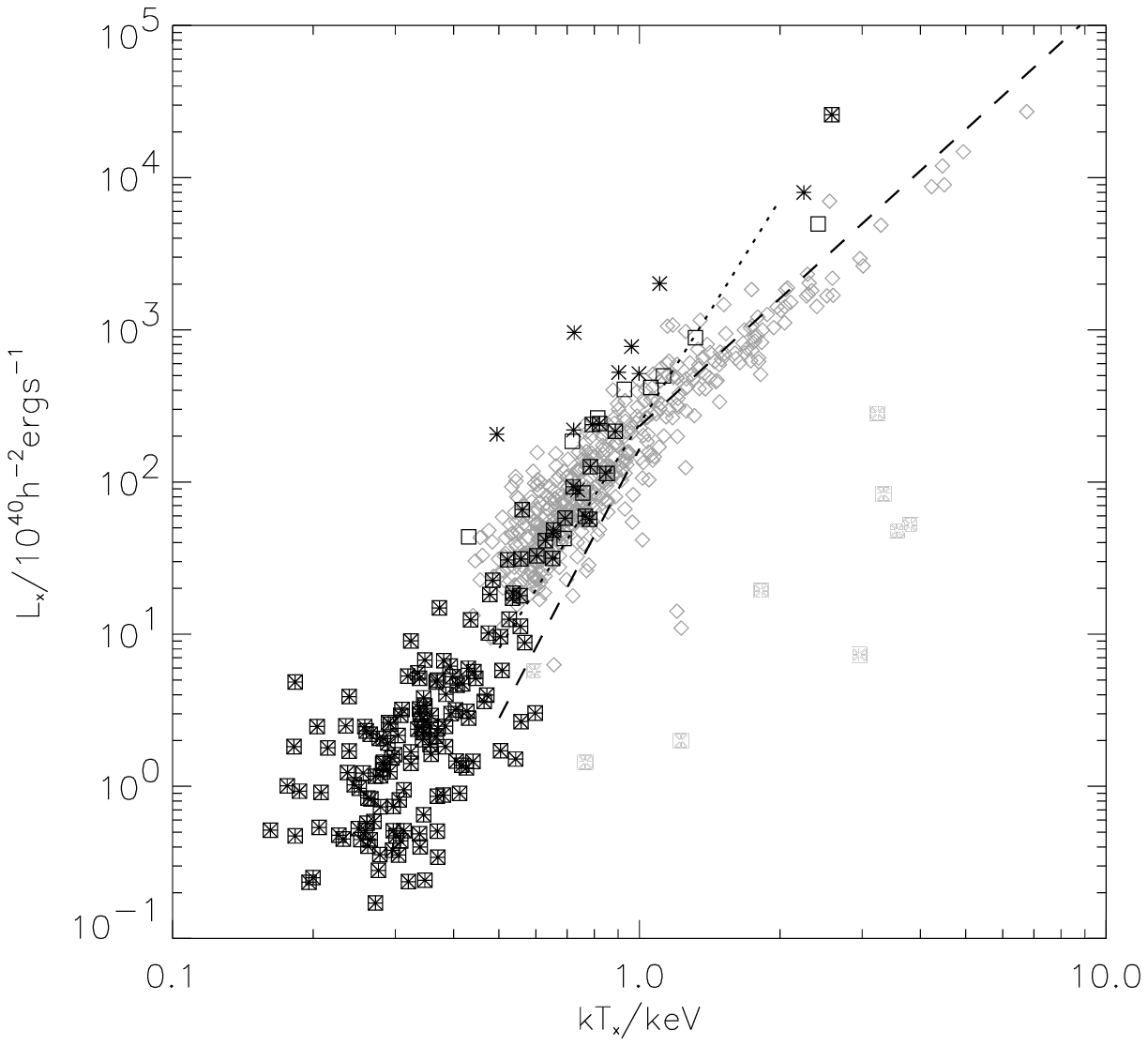,height=8cm}}
\caption{Luminosity-temperature relation for groups in the 
fiducial model (stars). Square points are the result 
when emission from within the cooling radius is excluded. 
Diamonds are from the {\it Radiative} simulation studied by M2002.
Dashed lines represent the best-fit relations to the group and
cluster samples compiled by Xue \& Wu (2000). The dotted line
illustrates the best-fit power law to the sample of groups found
by Helsdon \& Ponman (2000).}
\label{fig:lxtx_128}
\end{figure}

\begin{figure}
\centering
\centerline{\psfig{file=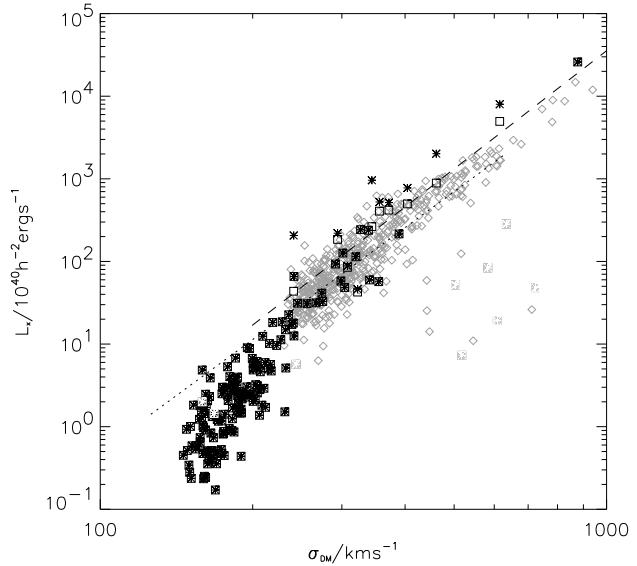,height=8cm}}
\caption{$L_{\rm x}-\sigma_{\rm DM}$ relation for groups in the 
fiducial model (stars). Square points are the result 
when emission from within the cooling radius is excluded. 
Diamonds are from the {\it Radiative} simulation studied by M2002.
Dashed lines represent the best-fit relations to the group and
cluster samples compiled by Xue \& Wu (2000). The dotted line
illustrates the best-fit power law to the sample of groups found
by Helsdon \& Ponman (2000).}
\label{fig:lxsigma_128}
\end{figure}

Figs.~\ref{fig:lxtx_128} \& \ref{fig:lxsigma_128} illustrate the 
$L_{\rm x}-T_{\rm x}$ and $L_{\rm x}-\sigma$ relations respectively,
for the fiducial simulation. Again, the model agrees reasonably well
with both M2002's result and the observed relations, given the uncertainty
in the latter. Furthermore, increasing the resolution of the smaller
objects (which are less well-resolved than the larger groups) increases
their temperature, which will bring the $L_{\rm x}-T_{\rm x}$ relation
into better agreement with the observed relation 
(see Section~\ref{sec:resolution}). Note that the ``break'' in the relation at
around 1 keV is more prominent in the $L_{\rm x}-T_{\rm x}$ relation 
than in the $L_{\rm x}-\sigma$ relation, due to the additional effect of entropy 
increase on temperature, causing the relation to steepen further. 

\subsection{Baryon fractions}

\begin{figure}
\centering
\centerline{\psfig{file=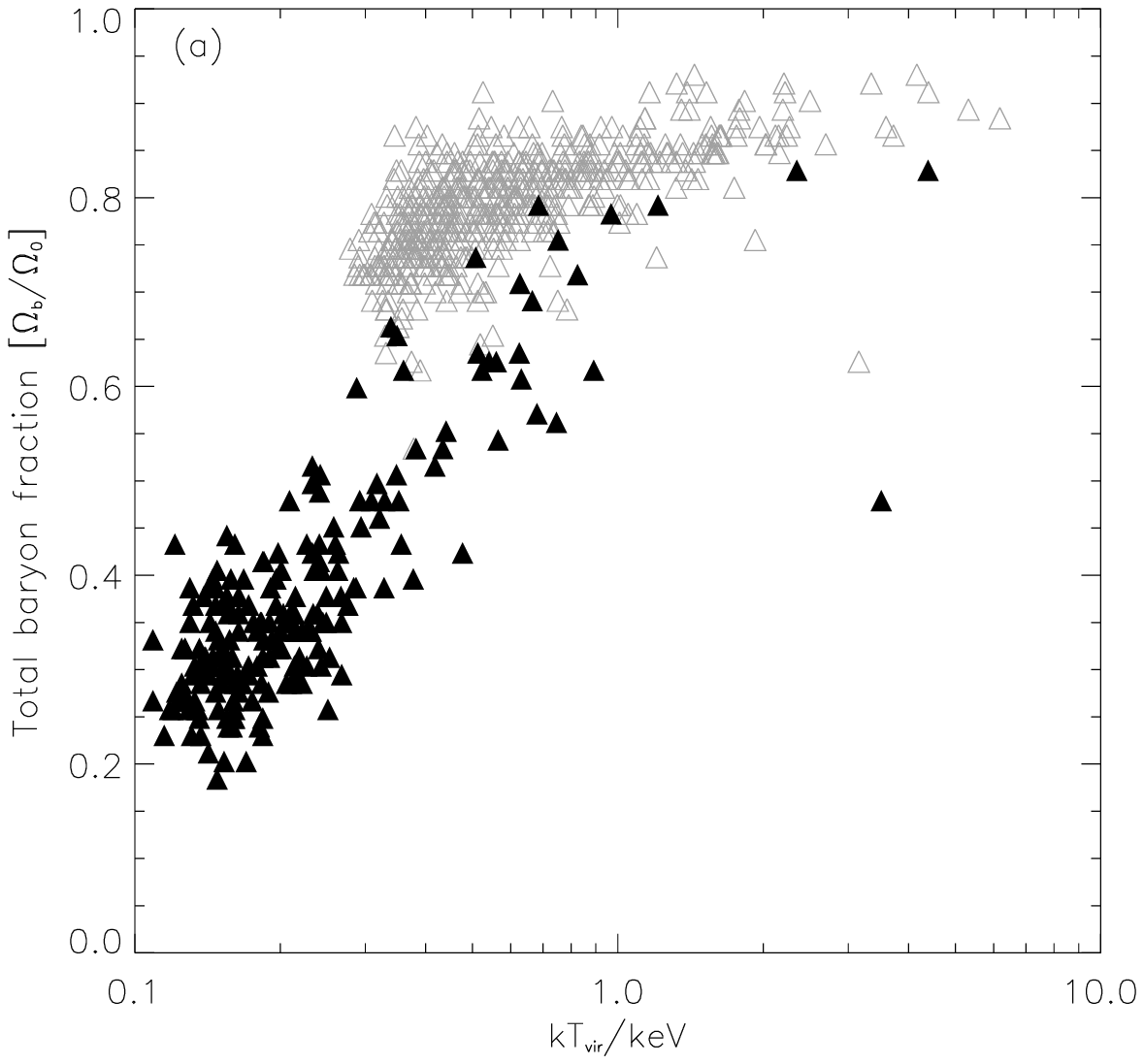,height=8cm}}
\centerline{\psfig{file=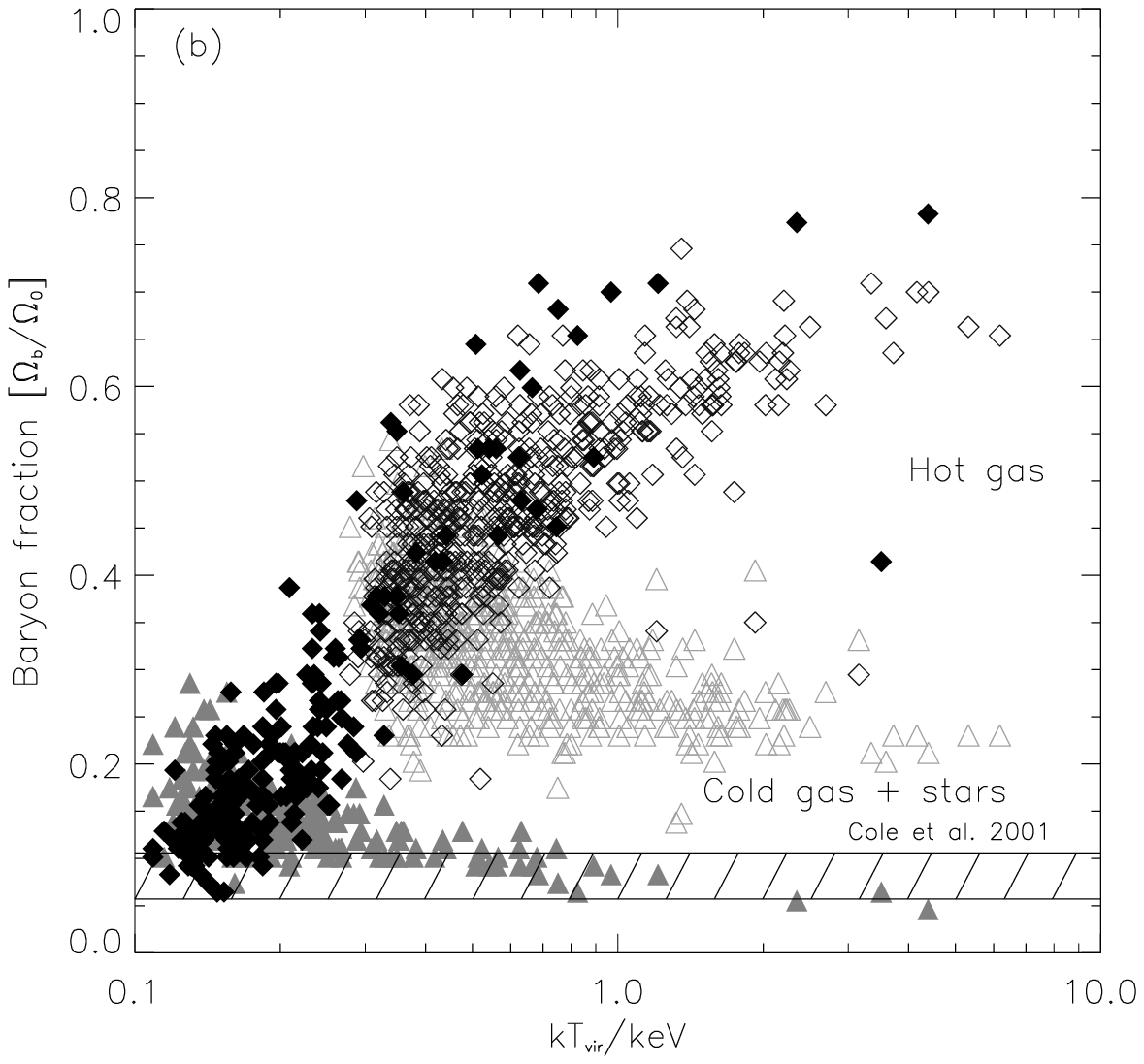,height=8cm}}

\caption{Baryon fractions, in units of the
global value, against virial temperature for groups 
in the fiducial simulation (filled symbols) and in the 
{\it Radiative} simulation studied by M2002 (open symbols). 
Total baryon fractions are plotted in panel (a), separated
into hot gas (diamonds) and cold gas + stars (triangles)
in panel (b). The shaded region represents the 2 global 
cooled fractions measured by Cole et al. (2001) for the 
Salpeter (upper horizontal line) and Kennicutt (lower line) IMF.}
\label{fig:bfrac}
\end{figure}

Although cooling and heating (feedback) are both mechanisms for increasing
the core entropy of groups, they have opposing effects on 
the stellar baryon fraction. Cooling acts to increase and feedback
decrease galaxy masses, hence matching the observed value requires a
balance between the two processes. Cooling and feedback can also have 
different effects on the hot gas fractions. Although cooling reduces the 
hot gas content, feedback can cause an increase (reheated gas that has not 
moved beyond the virial radius) or a decrease (transfer of energy 
between outflows and hot halo gas increases the fraction expelled from the 
halo).

In Fig.~\ref{fig:bfrac} we plot mass fractions of baryons, 
in units of the global value ($\Omega_{\rm b}/\Omega_{\rm m}$), 
against the virial temperature of each halo for the fiducial
model (filled symbols) and M2002's {\it Radiative} simulation 
(open symbols). The total baryon fractions are plotted in panel (a).
Both models appear to be converging towards a value close to unity at 
high virial temperatures, but the baryon fractions decrease much more 
rapidly at lower temperatures when feedback is included. The difference
between baryon fractions reflects the total amount of gas expelled from
the halo by feedback.

The baryon fractions are split into hot gas ($T>10^{5}$K, diamonds) 
and cold gas + star (triangles) fractions in panel (b). The
same trend is present between the two models for the cooled baryon
fractions (since galaxy formation efficiency is higher in smaller
objects, where average cooling times are shorter), but the
values at a given temperature are different. The cooled fraction
is twice as high in the M2002 objects, and could be higher: the 
absence of feedback in this simulation means that only numerical 
resolution limited these values. Feedback also steepens the relation 
between the hot gas fraction and virial temperature. As
we discuss in Section~4, objects with virial temperatures above the 
feedback temperature, $kT_{\rm SN}$, retain a fraction of reheated gas 
which slightly increases their hot gas fraction. In smaller systems however, 
feedback has a larger impact on the hot gas, expelling a significant
fraction from the halo.

The shaded region in Fig.~\ref{fig:bfrac} represents measured global 
cooled gas fractions (the global value is the asymptotic limit for 
large halo mass), combining the stellar fractions measured by Cole et al. 
using data from the 2MASS and 2dF galaxy surveys, with a further 10 per 
cent of the mass in cold gas (Balogh et al. 2001). The upper horizontal line 
represents the value determined when using the Salpeter IMF 
(as used in our feedback model) and the lower for the Kennicutt IMF.
Globally, only just over 3 per cent of the gas forms stars in the 
fiducial model by $z=0$, lower than the observed value by a factor of
$\sim 3$. Increasing the resolution does not significantly increase
the cooled fraction (Section~\ref{sec:resolution}) because feedback
acts to stabilize this quantity and our simulation has already resolved 
the majority of the galaxy mass density. As we discuss in the next section 
however, a higher cooled fraction can be achieved by choosing a lower 
$\epsilon$ which does not significantly affect the scaling relations. 
Additionally, increasing the metallicity of cooling gas will increase
the fraction of baryons in stars. The metallicity of high-density
regions has been underestimated in our simulation since stars in galaxies 
at the present day have been enriched to solar abundance 
(Edmunds \& Phillips 1997), whereas ours have $Z=0.3Z_{\odot}$. A higher 
metallicity will increase the fraction of stars in a given galaxy.

Measuring cooled fractions in groups and clusters is difficult, 
mainly due to the lack of sensitivity in the X-ray instruments
to measure hot gas masses out to the virial radius. Current attempts,
however (see for example the compilation of data by Balogh et al.), 
measure values between 10-20 per cent at 1 keV. These values are 
marginally consistent with our fiducial result but again, a lower
$\epsilon$ and/or higher metallicity would improve the match.

\section{A closer look at the model}
\label{sec:model}

\begin{table}
\caption{Simulations used to study the effects of feedback.
From left to right, each column gives the model studied,
feedback efficiency, reheating temperature, global fraction 
of baryons in stars at $z=0$ and whether the model provides a 
reasonable match to the $L_{\rm x}-T_{\rm x}$ relation.}
\begin{center}
\begin{tabular}{lllll}
\hline
Model & $\epsilon$ & $kT_{\rm SN}$/keV & $f_{\rm star}$ & 
$L_{\rm x}-T_{\rm x}$?\\
\hline
Radiative & N/A & N/A & 0.15 & YES\\
Feedback  & 1   & 0.1 & 0.10 & NO\\
Feedback  & 0.1 & 2   & 0.12 & YES\\
Feedback  & 1   & 2   & 0.05 & YES\\
Feedback  & 2   & 2   & 0.03 & YES\\
\hline
\end{tabular}
\label{tab:fb}
\end{center}
\end{table}

Given the ability of our fiducial model to adequately reproduce
many of the key X-ray observations of groups, we now investigate
how these results depend on the strength of the feedback.
To facilitate this task we have run a set of smaller simulations
(Table~\ref{tab:fb})
with $N=2\times 64^3$ particles within a comoving box of length
$25 \hMpc$, thus keeping the resolution constant. 

After generating group catalogues for each simulation, we also compare
the properties of the haloes in two models on an object-by-object basis. 
To do this we first found the closest pair, then rejected those which
had masses differing by more than 15 per cent or separations
greater than 20 per cent of the average virial radius. These criteria
find a good match for the majority of the haloes in the two catalogues.

\subsection{Cooling versus heating}

The first comparison we make is between the fiducial model
(which has strong feedback) and a run with no feedback at all 
(radiative model). This allows us to quantify the relative 
effects of heating and cooling on the X-ray properties of the 
groups. In particular, we are interested in knowing if the 
model behaves like that proposed by Voit \& Bryan (2001), i.e. 
cooling generates the entropy excess by removing hot gas, which 
then either forms stars or is reheated and transported to larger radii.
If the fiducial model does behave in this way, we expect the feedback only 
to reduce the fraction of galactic material, leaving the hot gas fraction
(and hence X-ray properties) intact.

\begin{figure}
\centering
\centerline{\psfig{file=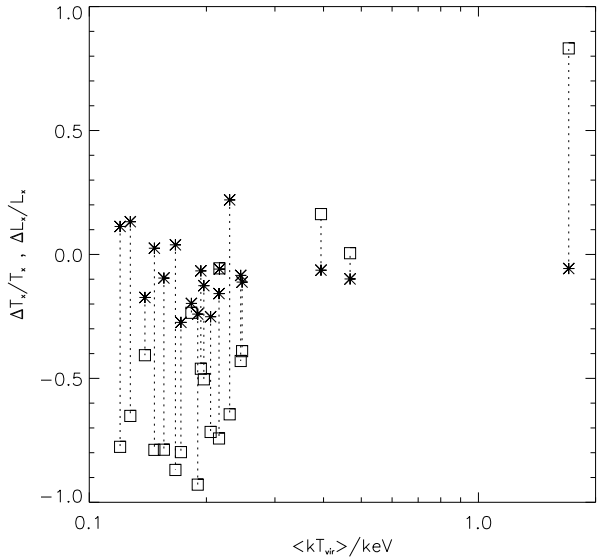,height=8cm}}
\caption{Fractional difference in X-ray temperature (stars)
and luminosity (squares) between matched haloes in the simulations 
with and without feedback, as a function of virial temperature.
Dotted lines are a guide to the eye.}
\label{fig:dx_cool}
\end{figure}

Indeed, the $L_{\rm x}-T_{\rm x}$ relation for the radiative simulation
is similar to the fiducial feedback relation, but the former is flatter.
This effect is quantified in Fig.~\ref{fig:dx_cool}, where we plot the fractional 
difference in X-ray temperature (stars) and luminosity (squares) between matched 
haloes in the two runs against their average virial temperature. These differences
are defined as
\begin{equation}
\Delta T_{\rm x}/T_{\rm x} = 1 - 
T_{\rm x}({\rm feedback})/T_{\rm x}({\rm radiative}),
\end{equation}
and
\begin{equation}
\Delta L_{\rm x}/L_{\rm x} = 1 - 
L_{\rm x}({\rm feedback})/L_{\rm x}({\rm radiative}).
\end{equation}
There is no strong variation in the difference between X-ray 
temperatures with system size, although there is a correlation in the 
X-ray luminosity difference. For $kT_{\rm vir} \sim 0.1$keV the objects 
in the radiative simulation are nearly twice as luminous as those in the 
feedback run, but the opposite is the case for systems which are a factor 
of 10 hotter. (Note that because of the small size of the simulations, 
there is only one halo with $kT_{\rm vir}>1$keV.) 

On closer inspection,
the trend in luminosity difference is down to two effects caused by feedback.
The first, which particularly affects low-mass systems, is a reduction
in the mass of hot gas in groups in the feedback simulation which was 
inside the virial radius in the radiative simulation. This can be
attributed to the increase in pressure from the outflowing reheated
material. Secondly, part of the reheated gas, which formed stars
in the radiative simulation, is still within the virial radius in the
feedback simulation. As the gravitational energy of the halo becomes comparable
to that given to reheated gas (i.e. $T_{\rm vir} \rightarrow T_{\rm SN}$), the
second effect dominates over the first, eventually causing the feedback
to increase the X-ray luminosity of a group.

To summarize, we find that our model does behave like
that proposed by Voit \& Bryan (2001) if $T_{\rm SN} \sim T_{\rm vir}$.
If $T_{\rm SN} \gg T_{\rm vir}$ then feedback also removes hot halo gas
(in addition to expelling reheated material) and so reduces the 
group's X-ray luminosity further. Conversely, if 
$T_{\rm SN} \ll T_{\rm vir}$, some reheated gas does not 
gain enough entropy to escape and so boosts the 
group's X-ray luminosity.

\subsection{Varying the model parameters}

We now investigate the effects of varying the feedback model 
parameters, $\epsilon$ and $T_{\rm SN}$ on the X-ray properties
of groups. Given the simplicity of the model, and the results from 
the previous comparison with a radiative simulation, we should be 
able to predict how varying the parameters affect the results.

\begin{figure}
\centering
\centerline{\psfig{file=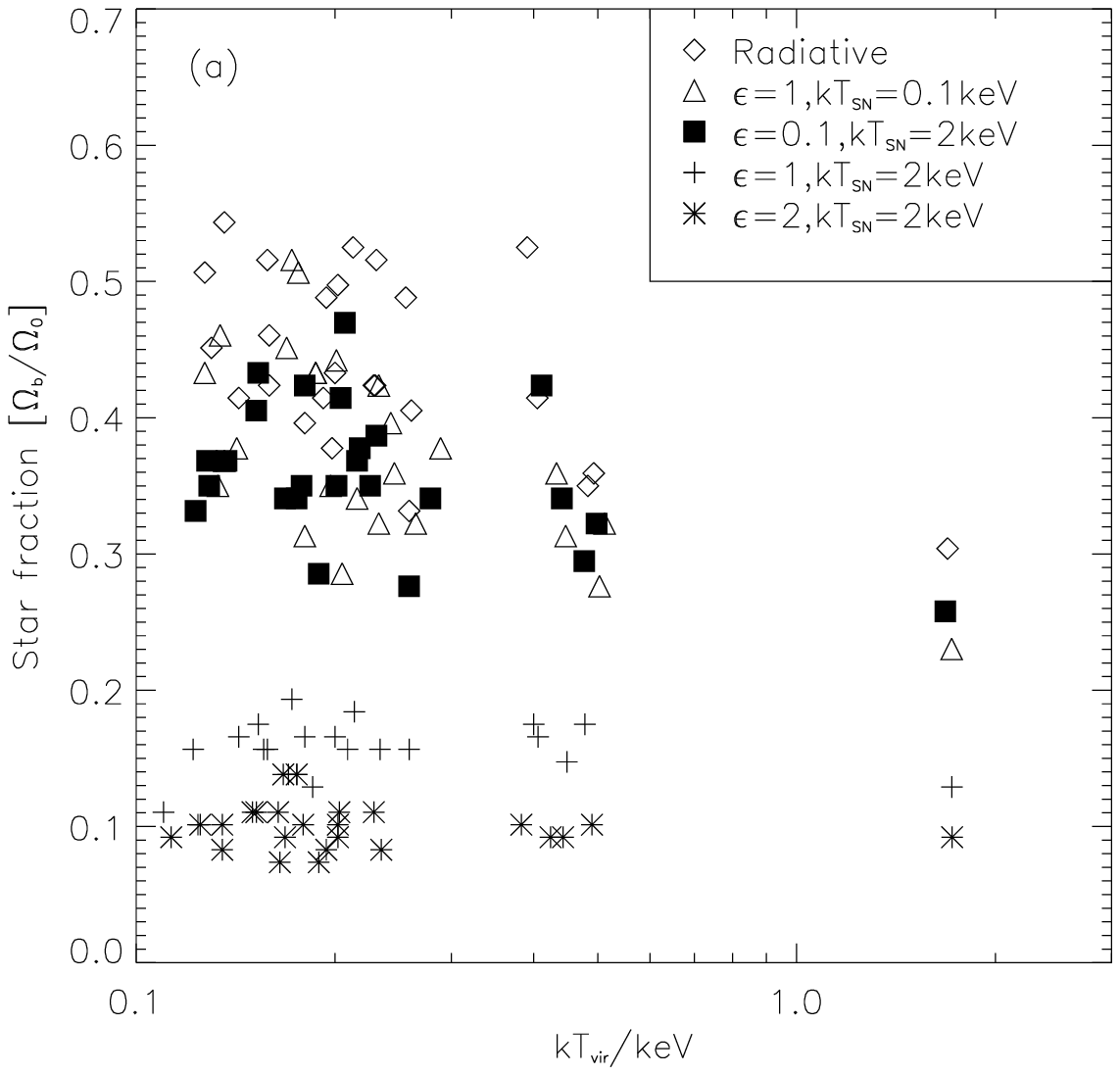,height=8cm}}
\centerline{\psfig{file=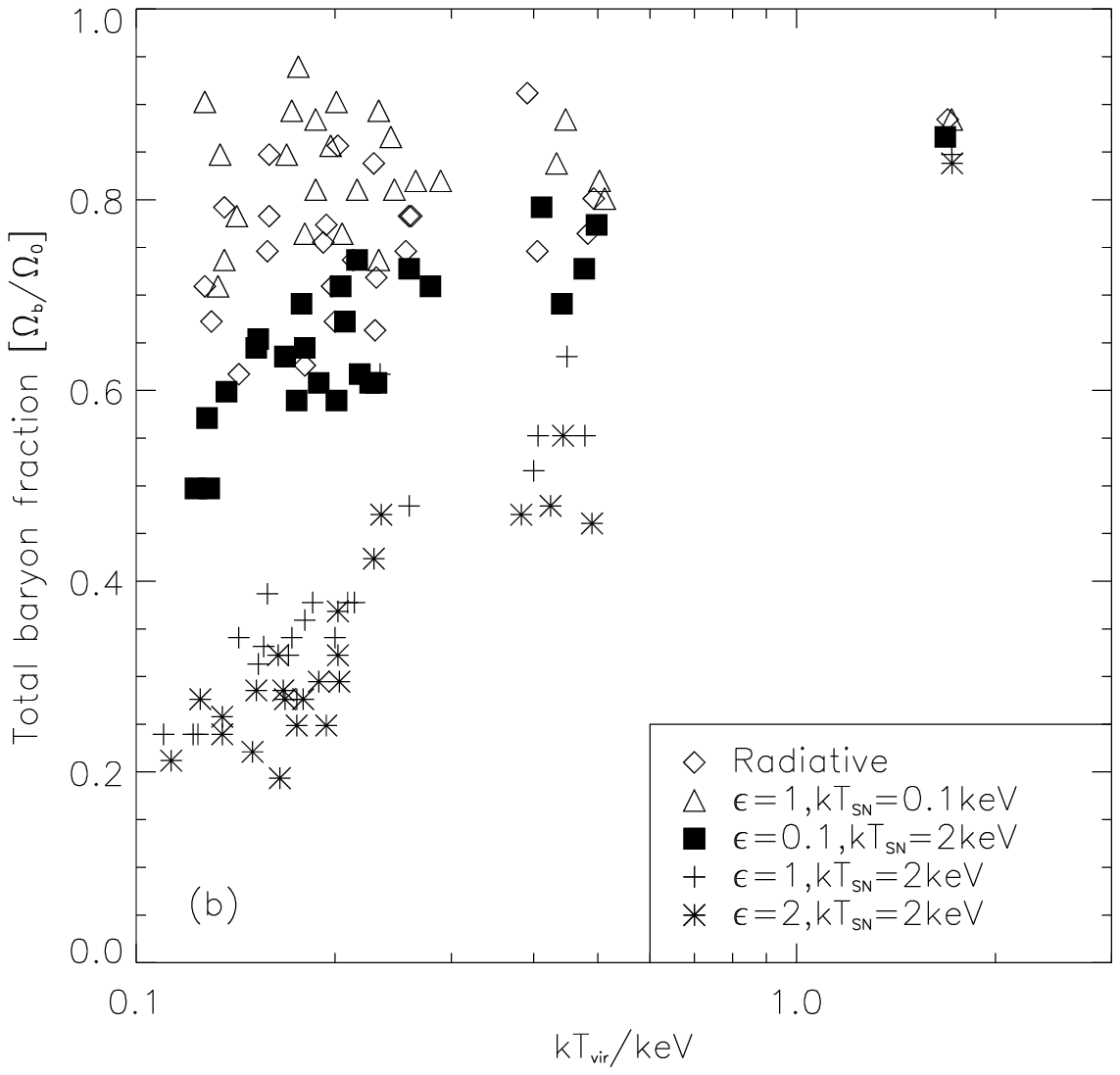,height=8cm}}
\caption{Fraction of baryons in stars (panel a)
and total baryon fraction (panel b) against 
halo virial temperature for the radiative run and
the runs with varying feedback parameters.}
\label{fig:starfrac}
\end{figure}

First of all, we consider the effect of varying the feedback parameters
on star formation efficiency. The star formation rate, which determines
the amount of energy and metals supplied to the interstellar medium,
varies inversely with $f_{\rm heat}$ (equation~\ref{eqn:fheat}), since 
more reheated cold gas leaves less material available to form stars. 
Note $f_{\rm heat} \propto \epsilon/kT_{\rm SN}$, so we would expect a 
smaller energy budget and/or a higher heating temperature to increase the
star formation rate.

Panel (a) of Fig.~\ref{fig:starfrac} illustrates star fractions against 
virial temperature for groups in the radiative model plus feedback models with 
varying parameter values (given in the legend). As expected, the star
fraction increases when decreasing $\epsilon$ for fixed $kT_{\rm SN}$,
and the model with $\epsilon=0.1$ contains approximately the correct
amount of cooled material ($\sim$ 10 per cent, c.f. Fig.~\ref{fig:bfrac}).
The impact of varying $\epsilon$ on the total baryon fraction is shown
in panel (b). As $\epsilon$ is increased, more gas is reheated (at
the expense of forming stars) and expelled from each halo, 
decreasing its baryon fraction. 

Contrary to expectations however, the star fraction increases when 
$kT_{\rm SN}$ is decreased from 2 to 0.1 keV, with the total baryon
fraction of the low temperature case being similar, on average, 
to the radiative model. As was found for K2002's thermal feedback model, 
if $T_{\rm SN}<T_{\rm vir}$ for the haloes being studied (as is the case 
for the $kT_{\rm SN}=0.1$ keV model), a substantial fraction of reheated 
gas is not hot enough 
\footnote{Note however that some gas will have been hot enough to escape from 
progenitor haloes at earlier times} 
to escape from the gravitational 
potential, and will instead re-cool and boost the star formation rate. 

\begin{figure}
\centering
\centerline{\psfig{file=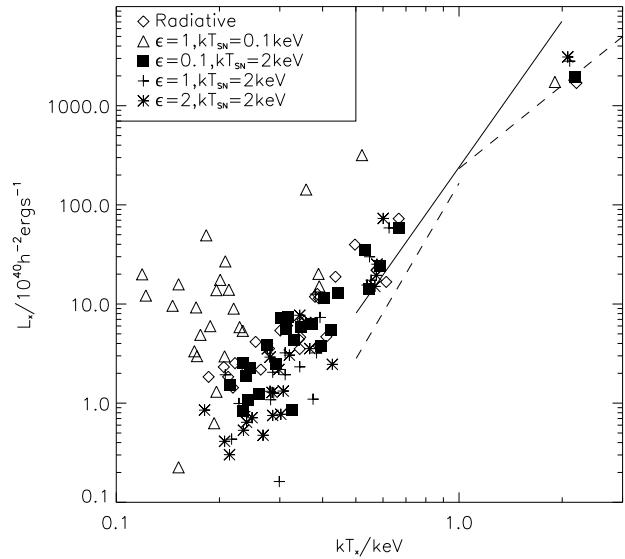,height=8cm}}
\caption{$L_{\rm x}-T_{\rm x}$ relations for the
radiative simulation and runs with varying feedback parameters.}
\label{fig:lxtx_64}
\end{figure}

The effect of feedback in the fiducial model was to slightly steepen the 
$L_{\rm x}-T_{\rm x}$ relation that was produced by radiative cooling
alone. Fig.~\ref{fig:lxtx_64} illustrates the effect of varying the 
parameters on the $L_{\rm x}-T_{\rm x}$ relation. Although the simulations
do not contain enough objects to perform a detailed study of the
shape of the $L_{\rm x}-T_{\rm x}$ relation, we see from our present 
results that the overall normalization of the relation is fairly insensitive to 
$\epsilon$ when $T_{\rm SN}>T_{\rm vir}$, except that the relation
becomes slightly steeper as $\epsilon$ is increased from 0 (radiative)
to 2. Although the heating temperature is the same in these runs, 
increasing $\epsilon$ increases the amount of outflowing material, thus
having a larger impact on the hot gas (as discussed in
the previous section). When $T_{\rm SN}<T_{\rm vir}$ however
(as is the case with $kT_{\rm SN}=0.1$keV) the luminosities are boosted 
and temperatures decreased considerably. As was observed from the 
previous figure, reheated material in the latter run is too cold to 
completely escape from a group-sized halo: the lack of buoyancy causes 
this gas to boost the central density, hence boosting its X-ray 
luminosity and lowering its X-ray temperature.

Although the fiducial run provides the best fit to the observations
(at least for low-mass groups), we will show in the next section that
the lack of numerical resolution in the smaller objects artificially
flattens the relation.

\section{Resolution effects}
\label{sec:resolution}

\begin{table}
\caption{Simulations used to study the effects of varying mass resolution}
\begin{center}
\begin{tabular}{ccccc}
\hline
Model & $N$ & $m_{\rm dark}/\hMsol$ & $\epsilon_{\rm p}/\hkpc$ & 
$f_{\rm cool}$\\ 
\hline
LR & $2\times 32^3$ & $5.2\times 10^{9}$  & $10$ & $0.031$ \\
HR & $2\times 64^3$ & $6.5\times 10^{8}$  & $5$  & $0.038$ \\
\hline
\end{tabular}
\label{tab:resol}
\end{center}
\end{table}

The results presented in this paper have all been for fixed
resolution. In this section, we present results from a limited
study of how varying the resolution affects the X-ray properties 
of the groups. Table~\ref{tab:resol} lists properties of the two 
simulations studied here, a low-resolution simulation (labelled LR) with
the same resolution used for the runs studied previously,
and a high-resolution simulation (HR) with 8 times as many
particles. Both runs were performed for a comoving box-size of 
$12.5 \hMpc$ and contain the same large-scale power. Although such a 
box is too small for quantitative predictions, it allows us to compare 
a small sample of the same objects at different resolution, without 
requiring a large investment in CPU time. The fiducial feedback
parameters were adopted, namely $\epsilon=2$ and $kT_{\rm SN}=2$keV.

The final column in Table~\ref{tab:resol} lists the cooled 
baryon fraction for the 2 runs; a factor of 8 increase in
resolution only increases this fraction by $\sim 20$ per cent.
This is because the majority of the mass in galaxies is already 
resolved at the lowest resolution (i.e. 32 or more particles, the
lowest number than can cool efficiently.). Furthermore, the 
feedback acts to stabilize gas cooling since the heating rate
is directly proportional to the cooling (star formation) rate.
As was stated previously, an overall increase in the cooled fraction 
must therefore come from choosing a lower value of $\epsilon$ and/or 
higher metallicity of cooling gas. 

\begin{figure}
\centering
\centerline{\psfig{file=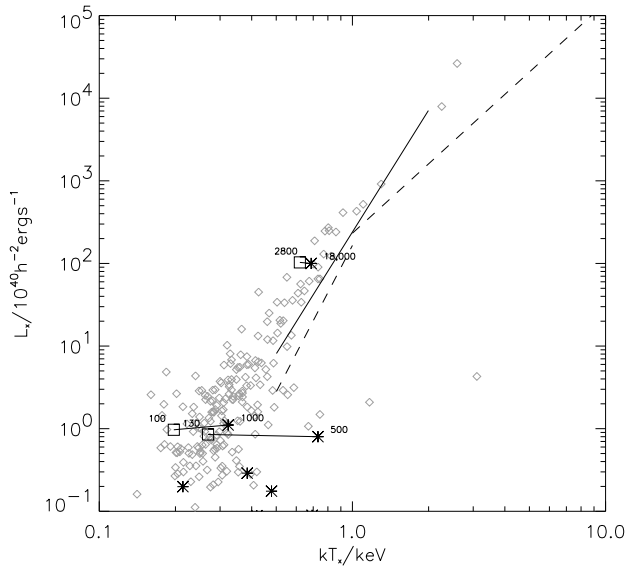,height=8cm}}
\caption{$L_{\rm x}-T_{\rm x}$ relation for the groups in run LR 
(squares) and HR (stars). Matched objects are connected with solid lines.
The $N=2\times 128^3$ fiducial relation (same resolution as LR) is also
plotted (diamonds). Numbers beside each point give the number of hot gas 
particles within the virial radius of the object. Solid and dashed 
lines illustrate best-fit observed relations presented by 
Helsdon \& Ponman (2000) and Xue \& Wu (2000) respectively.}
\label{fig:lxtx_resol}
\end{figure}

All runs resolve 3 common objects; the corresponding $L_{\rm x}-T_{\rm x}$ 
``relation'' for each simulation is plotted in Fig.~\ref{fig:lxtx_resol}. 
Particles from within the cooling radius of each object are excluded from
the calculation (which affects only the luminosity of the largest object). 
Note that the luminosities are more or less unchanged for all 3 objects 
when the resolution is varied, due to the stabilizing effect of feedback.
For the largest object, which contains between 2800 (LR) and 18,000 (HR)
hot gas particles, its X-ray temperature increases only a little, unlike
the smaller 2 objects (between 100 and 1000 particles), where it increases
significantly. We therefore 
expect the slope of the $L_{\rm x}-T_{\rm x}$ relation to steepen when 
the emission-weighted temperatures for all objects have converged. 

\section{Summary and conclusions}
\label{sec:conclusions}

In this paper, we examined the influence of the galaxy
population on the X-ray properties of groups at $z=0$ that form in
a cosmological simulation.  In particular, we were interested
in determining how including feedback of energy from supernovae in the
simulations (which decreases the star formation rate) affects the
X-ray properties of the groups.
Motivated by the ``break'' in the X-ray luminosity-temperature 
relation at the interface between group and cluster scales 
($kT_{\rm x} \sim 1-2$keV), we devised a feedback model based on heating 
the gas to a fixed temperature, independent of the available energy
budget. A given heating temperature, $kT_{\rm SN}$, and energy
budget, $\epsilon$, then determines the star formation rate and
the corresponding mass of reheated gas.

For our main results, we studied a model with strong feedback 
($\epsilon=2$, twice the available energy from supernovae) and a heating 
temperature, $kT_{\rm SN}=2$keV, comparable to the $L_{\rm x}-T_{\rm x}$ 
break. This model generated the required excess entropy in the cores of the 
groups and so was able to adequately reproduce the X-ray scaling relations 
over the range of temperatures appropriate to groups ($kT<2$keV). However, 
the fraction of baryons in galaxies was around a factor of 3 lower than 
observed (assuming a Salpeter IMF). 

We then compared our fiducial feedback model to a model with no
feedback, and found that the $L_{\rm x}-T_{\rm x}$ relations were similar 
in normalization, but the feedback relation was slightly steeper. 
Radiative cooling acts to reduce the hot gas fraction (and hence X-ray 
luminosity) in groups, removing low-entropy material (which forms stars) 
and causing higher-entropy gas to flow in to replace it. In low-mass groups 
(where $T_{\rm vir}<T_{\rm SN}$), feedback reduces the luminosity further 
because the outflowing reheated material increases the pressure of the
hot gas. However, as $T_{\rm vir} \rightarrow T_{\rm SN}$, 
the effect of feedback on the hot gas diminishes due to the group's increasing 
ability to retain a larger fraction of material. As a result, some of the 
gas that was turned into stars in the radiative run is still within the 
virial radius in the feedback run, leading to an increase in the luminosities of 
high-mass groups.

The fraction of reheated material per mass of stars scales as
$\epsilon/kT_{\rm SN}$, and so the star formation rate ought to
vary inversely with this parameter combination, since more reheated
gas leaves less material available to form stars. This is only true if the
heating temperature is comparable to or larger than the virial
temperature of the systems being studied. For such a case, decreasing
$\epsilon$ allows a reasonable match to both the global fraction of 
baryons in stars and the X-ray properties of the groups. However, when 
the heating temperature is low compared to the virial temperature of groups
(we studied the case when $kT_{\rm SN}=0.1$keV), reheated gas cannot escape 
from the gravitational potential and instead re-cools and forms more
stars. As a result, this excess of cooling gas decreases the core entropy of the
groups and consequently boosts their luminosity. 

Finally, increasing the resolution of our simulations appears to have 
little effect on the X-ray luminosities of groups but the temperatures of 
the smaller, poorly-resolved objects increased. Based on these results,
numerical convergence for the range of group masses will act to steepen the 
$L_{\rm x}-T_{\rm x}$ relation. 

\section*{Acknowledgments}
The work presented in this paper was carried out as part of the programme
of the Virgo Supercomputing Consortium 
({\tt http://virgo.sussex.ac.uk}). 
The majority of data were generated using COSMOS, the UK Computational
Cosmology Consortium supercomputer, an SGI Origin 3800 at the Department
of Applied Mathematics and Theoretical Physics, Cambridge.
STK is supported by PPARC; PAT and TT thank PPARC for a Lecturer Fellowship  
and Advanced Fellowship respectively.


\end{document}